\documentclass[10pt,journal,compsoc]{IEEEtran}
\usepackage[english]{babel}
\usepackage{hyperref}
\addto\extrasenglish{%
}

\usepackage{hyphenat}
\usepackage{xstring}
\usepackage{forloop}

\usepackage[defaultcolor=black]{changes}

\usepackage{enumitem}

\newsavebox\MyBreakChar%
\sbox\MyBreakChar{}
\newsavebox\MySpaceBreakChar%
\sbox\MySpaceBreakChar{\hyp}
\makeatletter%
\newcommand*{\BreakableChar}[1][\MyBreakChar]{%
  \leavevmode%
  \prw@zbreak%
  \discretionary{\usebox#1}{}{}%
  \prw@zbreak%
}%
\makeatother

\newcounter{index}%
\newcommand{\AddBreakableChars}[1]{%
  \StrLen{#1 }[\stringLength]%
  \forloop[1]{index}{1}{\value{index}<\stringLength}{%
    \StrChar{#1}{\value{index}}[\currentLetter]%
    \IfStrEqCase{\currentLetter}{%
        {*}{\currentLetter\BreakableChar[\MyBreakChar]}%
        {/}{\currentLetter\BreakableChar[\MyBreakChar]}%
        {+}{\currentLetter\BreakableChar[\MyBreakChar]}%
        {\&}{\currentLetter\BreakableChar[\MyBreakChar]}%
    }[\currentLetter]%
  }%
}%

\def\fEnvironment{\textsc{Environment}}
\def\fInteraction{\textsc{Interaction}}

\DeclareRobustCommand{\inlinefigs}[1]{%
\begingroup
\setbox0=\hbox{\includegraphics[height=1em]{#1}}%
\parbox[c][11pt][t]{\wd0}{\box0}\endgroup
}
\newcommand{\fVR}{\inlinefigs{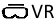}}
\newcommand{\fDesktop}{\inlinefigs{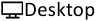}}
\newcommand{\fWIMP}{\inlinefigs{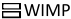}}
\newcommand{\fGesture}{\inlinefigs{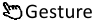}}

\newcommand{\vrWIMP}{\fVR{}+\BreakableChar{}\fWIMP{}}
\newcommand{\vrGesture}{\fVR{}+\BreakableChar{}\fGesture{}}
\newcommand{\desktopWIMP}{\fDesktop{}+\BreakableChar{}\fWIMP{}}
\newcommand{\desktopGesture}{\fDesktop{}+\BreakableChar{}\fGesture{}}

\newcommand{\fVRBold}{\inlinefigs{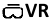}}
\newcommand{\fDesktopBold}{\inlinefigs{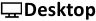}}
\newcommand{\fWIMPBold}{\inlinefigs{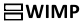}}
\newcommand{\fGestureBold}{\inlinefigs{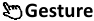}}

\newcommand{\vrWIMPBold}{\fVRBold{}+\BreakableChar{}\fWIMPBold{}}
\newcommand{\vrGestureBold}{\fVRBold{}+\BreakableChar{}\fGestureBold{}}
\newcommand{\desktopWIMPBold}{\fDesktopBold{}+\BreakableChar{}\fWIMPBold{}}
\newcommand{\desktopGestureBold}{\fDesktopBold{}+\BreakableChar{}\fGestureBold{}}

\ifCLASSOPTIONcompsoc
  \usepackage[nocompress]{cite}
\else
  \usepackage{cite}
\fi
\ifCLASSINFOpdf
\else
\fi
\begin{document}
%
\title{This is the Table I Want! Interactive Data Transformation on Desktop and in Virtual Reality}
%
%
%
%

\author{Sungwon In, Tica Lin, Chris North, Hanspeter Pfister,~\IEEEmembership{Fellow,~IEEE,} and Yalong Yang
\IEEEcompsocitemizethanks{
\IEEEcompsocthanksitem S. In and C. North are with the Department of Computer Science, Virginia Tech, Blacksburg, VA, 24060. E-mail: \{sungwoni, chnorth1\}@vt.edu \protect
\IEEEcompsocthanksitem Y. Yang is with the School of Interactive Computing, Georgia Institute of Technology, Atlanta, GA, 30309, and was with the Department of Computer Science, Virginia Tech. \protect 
E-mail: yalong.yang@gatech.edu
\IEEEcompsocthanksitem T. Lin and H. Pfister are with John A. Paulson School of Engineering and Applied Sciences, Harvard University, Cambridge, MA, 02138. E-mail: \{mlin, pfister\}@g.harvard.edu
\IEEEcompsocthanksitem Y. Yang is the corresponding author.
}
\thanks{Manuscript received April 19, 2005; revised August 26, 2015.}}

\markboth{Journal of \LaTeX\ Class Files,~Vol.~14, No.~8, August~2015}%
{Shell \MakeLowercase{\textit{et al.}}: Bare Demo of IEEEtran.cls for Computer Society Journals}
%




\newcommand{\insertfig}{\includegraphics[width=\textwidth, height=100pt]{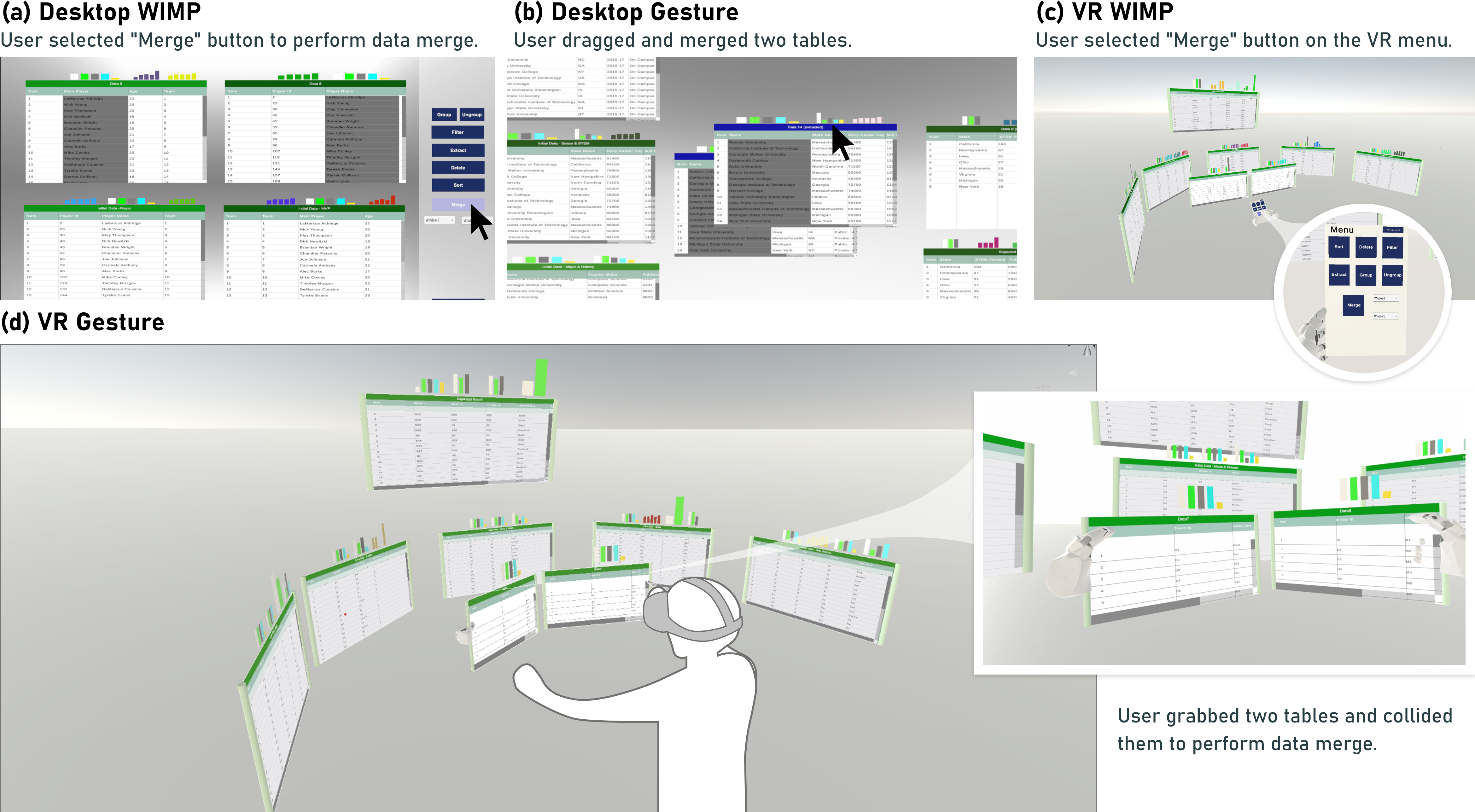}\captionof*{figure}{Four conditions designed for performing data transformation in the user study, including a combination of desktop or VR environments, and WIMP or gesture interactions.}\label{fig:my_label}}

\IEEEtitleabstractindextext{%
\begin{abstract}
Data transformation is an essential step in data science. While experts primarily use programming to transform their data, there is an increasing need to support non-programmers with user interface-based tools. 
With the rapid development in interaction techniques and computing environments, we report our empirical findings about the effects of interaction techniques and environments on performing data transformation tasks. Specifically, we studied the potential benefits of direct interaction and virtual reality (VR) for data transformation. 
We compared gesture interaction versus a standard WIMP user interface, each on the desktop and in VR. 
With the tested data and tasks, we found time performance was similar between desktop and VR.
Meanwhile, VR demonstrates preliminary evidence to better support provenance and sense-making throughout the data transformation process.
Our exploration of performing data transformation in VR also provides initial affirmation for enabling an iterative and fully immersive data science workflow.
\end{abstract}

\begin{IEEEkeywords}
Immersive Analytics, Data Transformation, Data Science, Interaction, Empirical Study, Virtual/Augmented/Mixed Reality
\end{IEEEkeywords}}

\maketitle

\section{Introduction}
\label{sec:intro}

Data transformation is a data science process that converts a data set into the desired format to enable subsequent data science tasks, like visualization and modeling~\cite{wickham2016r}.
It is well recognized that data scientists need to spend an excessive amount of time doing data transformation, making it essential but also the most tedious and time-consuming aspect of a data science project~\cite{kandel_wrangler_2011,guo2011proactive}.
Using a programming language, like SAS, R, or Python, is the standard way of performing data transformation.
However, as data science becomes ubiquitous and exposed to people with limited programming knowledge, the prerequisite of knowing to program makes data science inaccessible to a large group of professionals whose workflows involve data~\cite{kandel2011research}, which we called \textit{non-technical data workers}.
As a result, the back-and-forth communications caused by data science's iterative and open-ended nature can heavily inhibit insight discovery and decision-making.

In response, like in many other data science processes, there is an increasing trend of providing user interface based (UI-based for short) tools for data transformation (e.g., Tableau~\cite{tableauDesktop} for visualization and AutoML~\cite{amazon,google} for modeling).
These UI-based tools lower the entry barrier for data science and also help reduce errors~\cite{wang_autods_2021}. 
However, even though there exist several commercial UI-based data transformation tools (e.g., Tableau Prep Builder~\cite{tableauPrep}, Trifacta~\cite{trifacta}, and Alteryx~\cite{alteryx}), the field lacks an empirical understanding of people's experiences in using these tools and the considerations in designing them.
Moreover, while the WIMP (windows, icons, menus, pointer) metaphor is typically used for constructing UI-based tools, modern interaction techniques that allow direct manipulation of the visual elements in the same space (named \emph{embedded interaction})~\cite{saket2017evaluating,saket2019investigating} were found to be more time-efficient in specific scenarios, like manipulating visualizations~\cite{sarvghad_embedded_2019,drucker_touchviz_2013}.
Most notably, Kandel et al.~\cite{kandel_wrangler_2011} and Nandi et al.~\cite{nandi2013gestural,nandi2013interactive,jiang2013gesturequery} found their UI-based tools to be more efficient in low-level data transformation tasks.
However, it is unclear if their findings can be generalized to more realistic and complicated scenarios.
\textbf{Our first goal} is to investigate the potential benefits of embedded interaction techniques over traditional WIMP interfaces in more realistic data transformation tasks.

On the other hand, in addition to interaction techniques, the rapidly evolved display and interaction environments (e.g., virtual and augmented reality or VR/AR) offer tremendous opportunities for creating innovative human-computer interaction experiences.
Specifically, there is a growing interest in using VR/AR for data analysis, bringing an emerging research topic~---~Immersive Analytics~\cite{ens2021grand,marriott2018immersive}.
From recent studies, there are two most frequently reported motivations for using VR/AR in analytics: \emph{large display space}~\cite{satriadi2020maps,kwon2016study,yang2020embodied,lisle_evaluating_2020} and \emph{embodied interaction}~\cite{yang2020tilt,cordeil2017imaxes,bach2017hologram}.
We believe there is great potential to explore whether those identified benefits can be generalized in improving the data transformation workflow.
On the other hand, standard mid-air methods in VR/AR are not suitable for tasks requiring high-precision interactions~\cite{drey2020vrsketchin,mendes2016benefits}, which could be inevitable in some data transformation tasks.
Therefore, \textbf{our second goal} is to investigate how these identified pros and cons can affect data transformation tasks in immersive environments.

Data science is iterative by its nature and does not follow a sequential pipeline. Consequently, alternating between different steps is inevitable~\cite{wickham2016r}. For example, after observing some visualizations, analysts may need to perform extra data transformations for the next analysis iteration.
Although visual exploration (e.g., ImAxes~\cite{cordeil2017imaxes} and DataHop~\cite{hayatpur2020datahop}) is feasible in a fully immersed manner, there is no immersive data transformation tool (i.e., tools for explicitly changing data table formats).
When analysts want to use immersive visualization, they have to switch between VR and desktop to complete the iterative data science tasks, causing high overhead for context-switching.
To this end, we study immersive data transformation tools to progress to a future where analysts can be fully immersed in VR for the entire data science workflow and maximize the benefits of the next generation of display and interaction environment.

\begin{figure*}
    \centering
    \includegraphics[width=\textwidth]{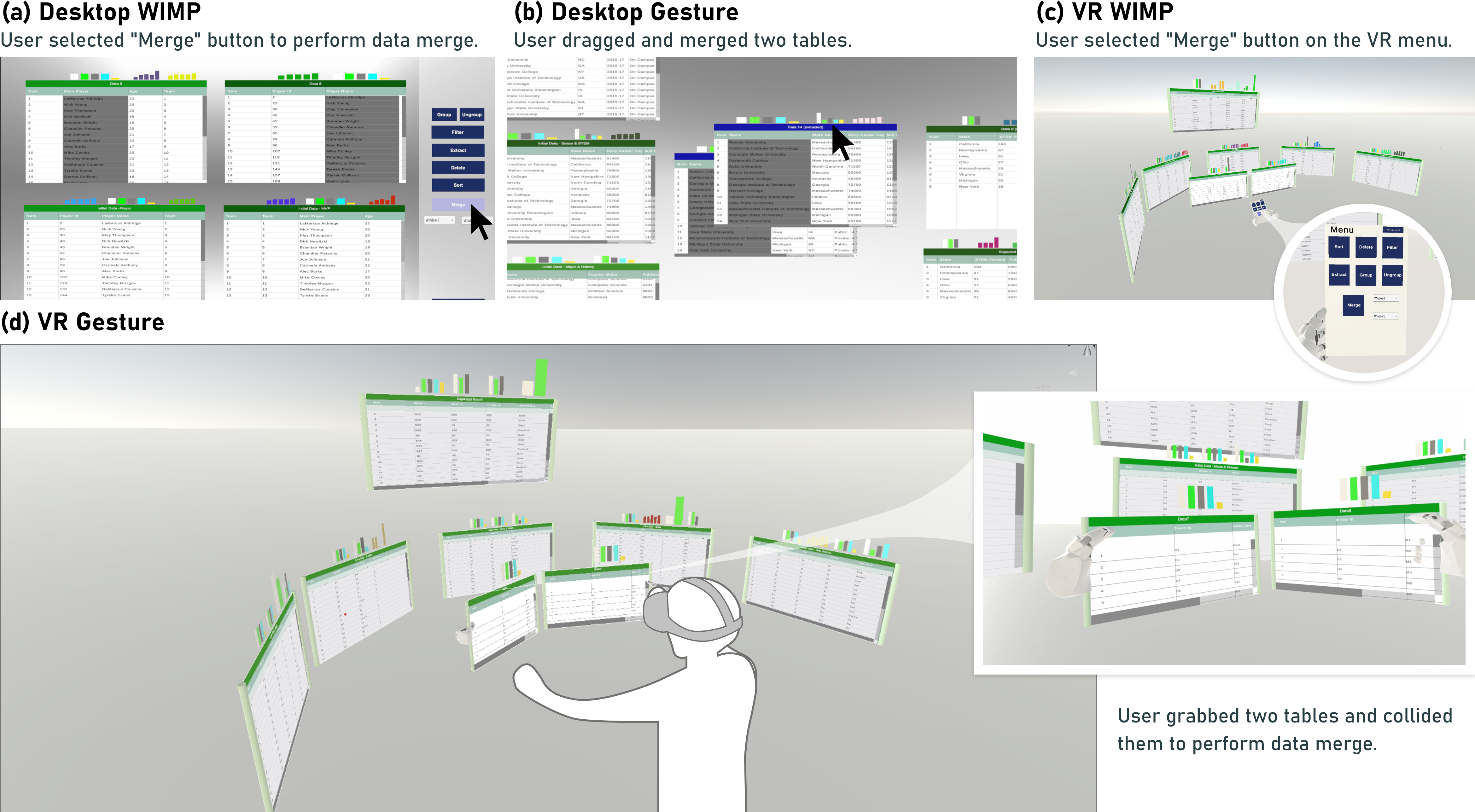}
    \vspace{-7mm}
    \caption{Four conditions designed for performing data transformation in the user study, including a combination of desktop or VR environments, and WIMP or gesture interactions.}
    \label{fig:teaser} 
    \vspace{-4mm}
\end{figure*}

To close these gaps, we developed prototypes with embedded interactions on the desktop and embodied interactions in VR for non-technical data workers to support essential data transformation operations.
Compared to a standard WIMP user interface, users can directly manipulate data tables through mouse or physical gestures (e.g., overlay one table on top of another to \emph{merge} them, see Fig.~\ref{fig:teaser}).
We compared our interaction designs to WIMP for desktop and VR.
To best simulate real-world scenarios, instead of testing low-level tasks, we asked participants to transform a set of data tables into a target format.
We found that participants required a similar amount of time to complete data transformation in VR and on a desktop. 
Meantime, VR demonstrated the potential to facilitate strategic thinking and support provenance better.
Subjectively, participants found the WIMP user interface on a desktop most familiar, and using VR was more physically demanding. 
On the positive side, VR was perceived as more engaging, and participants overall preferred the gesture-based experience in VR.
\textbf{The contributions of this paper are twofold}: 
\textit{first}, the designs of gesture-based interactions for essential data transformation operations on desktop and in VR; 
\textit{and second}, a user study systematically investigating the effect of interaction methods (WIMP vs. gesture) and computing environments (desktop vs. VR) on performing essential data transformation operations.

\vspace{-1em}
\section{Related Work}
\label{sec:related_work}

Our work is built upon two lines of prior work: data transformation tools and interaction methods. We also extend immersive analytics by enabling data transformation, an essential data science workflow, in immersive environments.

\vspace{-1em}
\subsection{Data Transformation Tools}

\textbf{Programming-based tools} are widely used by people with expertise and experience in programming. A wide range of libraries has been developed to support data transformation, like Pandas~\cite{mckinney2010data}, dplyr~\cite{dplyr}, tidyr~\cite{tidyr}, and plyr~\cite{tidyr}.
Programming-based tools are \emph{expressive}, and users can use their almost ``exhaustive'' APIs and parameters to complete various data transformation tasks.
However, mastering these tools requires extensive training and leads to a steep learning curve. Debugging complicated scripts is also oftentimes challenging~\cite{xiong_visualizing_2022}.
There are a series of attempts to address these issues.
As representative examples, DataLore~\cite{datalore} provides code suggestions to speed up the data transformation process.
Along a similar line, Wrex~\cite{drosos2020wrex} uses the notion of programming-by-example to generate data transformation code. 
On the other hand, Somnus~\cite{xiong_visualizing_2022} visualize data transformation scripts to help debug and gain a better overview of the process.
Yet, users of these tools are still expected to be experienced in programming, which excludes a large group of non-technical data workers.

\textbf{UI-based tools} allow people to manipulate their data without programming knowledge. 
Microsoft Excel is undoubtedly the most popular UI-based data transformation tool. 
Performing simple operations (e.g., editing values, sorting and filtering a column) is straightforward with its WIMP user interface. However, more complicated operations (e.g., merging two tables) requires knowing the specific ``secret'' menu item or writing code.
A few commercial UI-based tools aim to allow the users to quickly find the needed menu items, like Tableau Prep Builder~\cite{tableauPrep}, Trifacta~\cite{trifacta}, and Alteryx~\cite{alteryx}.
Wrangler~\cite{kandel_wrangler_2011}, meanwhile, provides natural language as ways users can specify the intended operations. However, users could get into trouble with discoverability (i.e., the ability to find and execute features) as a common challenge faced by a conversational user interface~\cite{kirschthaler2020can}.
GestureDB project enables gestures to describe the intended database queries~\cite{nandi2013gestural,jiang2013gesturequery}.
Our gesture design on the desktop environment shares many similar characteristics with their system, and we adapted and extended the gesture-based data transformation method to VR.

\textbf{Empirical results.} Some of the proposed tools have been evaluated, for example, Wrex was found more beneficial than a standard programming interface~\cite{drosos2020wrex}, and Wrangler was found to outperform Excel~\cite{kandel_wrangler_2011}.
Additionally, the GestureDB system was found to be more effective than the programming interface and non-gesture UI-based interface in performing single operations~\cite{nandi2013gestural,jiang2013gesturequery}.
We focus on UI-based data transformation tools as they lower the barrier for non-technical data workers. 
Our study aims to enrich the empirical understanding of using UI-based data transformation tools in tasks that require a series of operations.

\subsection{Embedded and Embodied Interactions}
We consider both embedded and embodied interaction under the same notion of \emph{direct manipulation} of visual representations. Direct manipulation contrasts with the standard WIMP UI design, which requires users to trigger operations on a space-separated area different from the area with visual representations.

Performing direct manipulation on a flat screen is considered \textbf{embedded interaction}.
It has been widely used in many UIs. For example, when uploading an email attachment, instead of clicking a button to select a file from a newly opened file browser, people can drag\&drop the file into the window.
In addition to this simple example, it has been used in many other applications, like annotation~\cite{shneiderman2000direct}, image editing~\cite{jia2006drag}, and content organization~\cite{lekschas2020generic}.
For data science, some work explored embedded interaction in manipulating data visualization~\cite{saket2017evaluating,sarvghad_embedded_2019,drucker_touchviz_2013,rzeszotarski2014kinetica,pirolli1996table}, analysis~\cite{horak2020responsive,burley2019arquery} and modeling~\cite{crotty2015vizdom}.
Most relevant to our work, GestureDB demonstrated some benefits of embedded interaction for elementary operations~\cite{nandi2013gestural,jiang2013gesturequery}, and we aim to study its potential benefits in more complex data transformation tasks.

Performing direct manipulation using body movement is considered \textbf{embodied interaction}.
The ability to track physical movement is essential to enable embodied interaction, which is an intrinsic characteristic of VR. 
As a result, many basic VR interactions are embodied. For example, grab\&move virtual objects, and rotate the head to change the viewpoint.
Embodied interaction has been explored for authoring visualizations~\cite{cordeil2017imaxes}, navigating in space~\cite{yang2020embodied, andrews2010space}, and switching between different views~\cite{yang2020tilt}.
We are interested in how we can adapt and extend the gesture designs from desktop to VR and whether the benefits of embodied interaction can be generalized to data transformation tasks.

As evaluated in the aforementioned works, one of the motivations of \emph{direct manipulation} design is to reduce the number of context-switching needed by having the interaction and visual representation in the same display area. We are interested in if this identified benefit can facilitate data transformation tasks.

\subsection{Immersive Analytics Toolkits}
Immersive Analytics has exploded into a fast-growing body of research on techniques and toolkits~\cite{ens2021grand}.
Existing Immersive Analytics research strongly focuses on data visualization~\cite{fonnet2019survey,marriott2018immersive}.
Specifically, a few toolkits enable data scientists to create immersive data visualizations, including DXR~\cite{sicat2018dxr}, VRIA~\cite{butcher2020vria}, IATK~\cite{cordeil2019iatk}, DataHop~\cite{hayatpur2020datahop} and ImAxes~\cite{cordeil2017imaxes}.
While ImAxes and DataHop provide a fully immersive visualization authoring experience, the others require users to create and configure visualizations on the desktop and view visualizations in the immersive environment.
More importantly, a data science project is oftentimes iterative and includes more than just data visualization. 
Immersive data visualization alone cannot fully leverage immersive environments for data analysis.
In this study, we explore how we can enable immersive data transformation to make progress toward a fully immersive data science workflow.

\begin{figure*}
    \centering
    \includegraphics[width=0.8\textwidth]{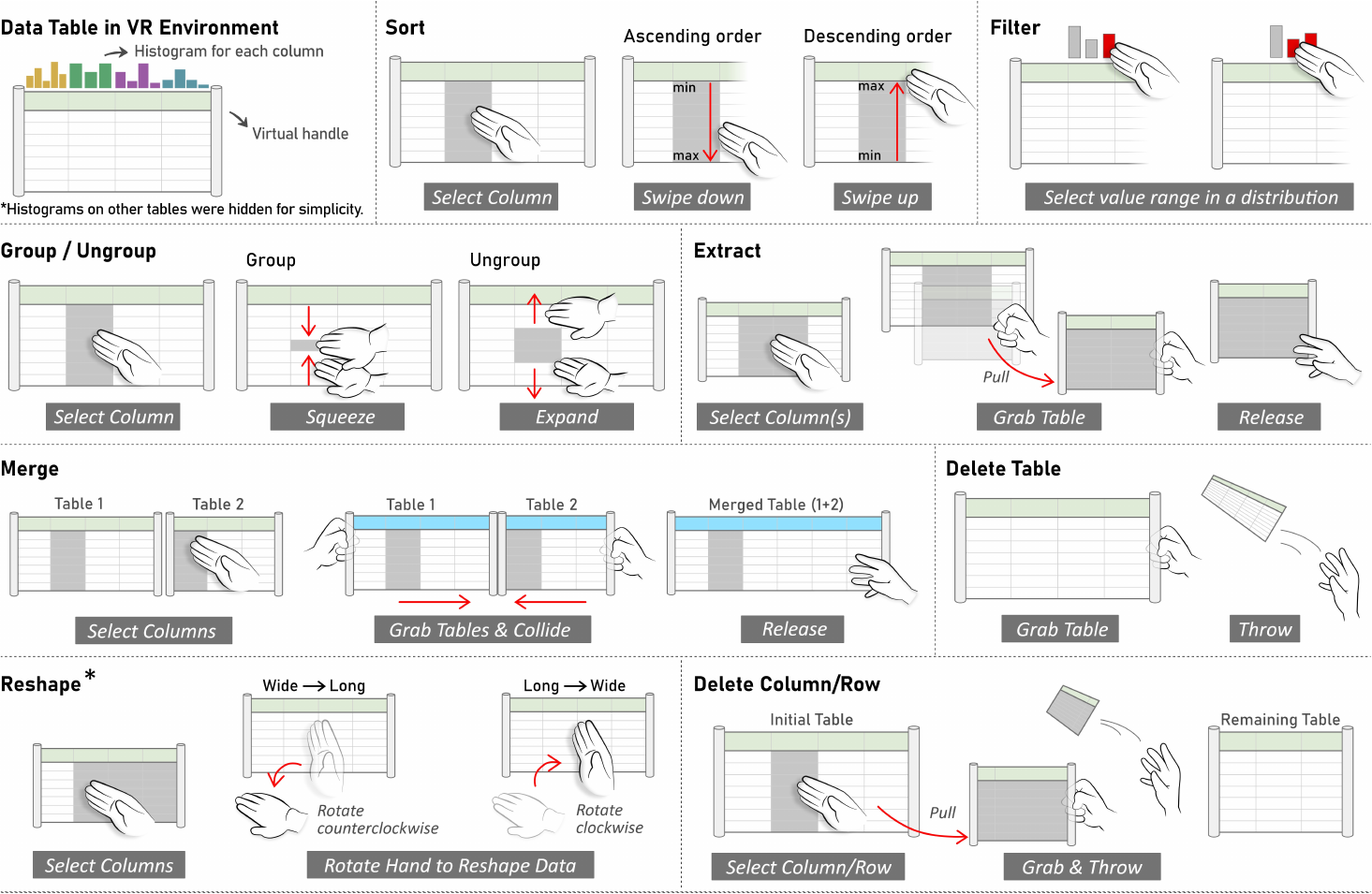}
    \vspace{-3mm}
    \caption{VR data table and gestures for data transformation tasks in the VR+Gesture condition. VR controllers, represented as hand models in the VR prototype and in the figure above, are used to perform these operations. The operation that was excluded from the study is marked with *.}
    \label{fig:vrgesture} 
    \vspace{-4mm}
\end{figure*}

\section{Embedded and Embodied Gesture Design}
Our study is intended to investigate the opportunities in using novel interaction methods (i.e., embedded and embodied Gesture) and emerging computing environments (i.e., VR) to better support data transformation.
We first reviewed the literature to identify the necessary operations and then designed gestures for both Desktop and VR.


\subsection{Selecting Data Transformation Operations}

Kasica et al.~\cite{kasica_table_2021} summarized 21 fine-grained data transformation operations across five categories (i.e., create, delete, transform, separate, and combine) on three targets (tables, columns, and rows), see Table~\ref{table:operations}.
As the first attempt to compare data transformation experience on desktop and in VR, we focus on basic operations that are commonly used by non-technical data workers without the need for programming.
To this end, we excluded operations that require programming-like input, including \textit{create}, \textit{transform}, \textit{separate}, and \textit{combine} operations on rows and columns. For example, one typical \textit{combine columns} operation can be inputting a formula to calculate the weighted average of selected columns.
One exception was the \textit{summarize rows} operation, as it does not require inputting a formula.

In summary, we support 12 operations, including all nine table, one-column, and two-row operations. See Table~\ref{table:operations}.

\begin{table}
    \centering
    \caption{A summary of our supported data transformation operations and their matching gestures.}
    \vspace{-3mm}
    \includegraphics[width=0.5\columnwidth]{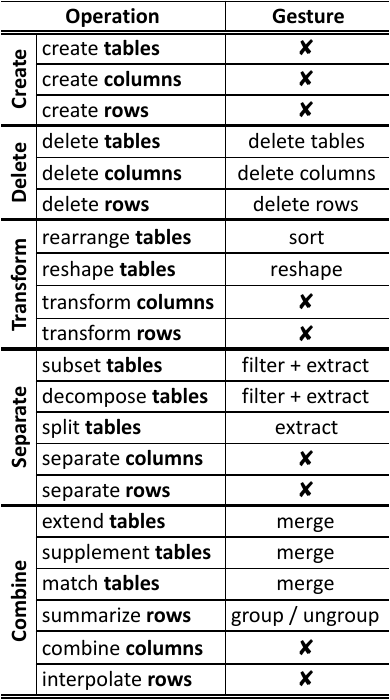}
    \label{table:operations}
    \vspace{-3mm}
\end{table}

\subsection{Designing Gestures}
\label{sec:operations}
After analyzing the selected operations, we found that some operations have differences at the semantic level but imply the same interaction analogy.
Specifically, Kasica et al. described \emph{extend}, \emph{supplement}, and \emph{match} as operations to combine two data tables, with their difference being row-wise vs. column-wise combination or inner-join vs. outer-join~\cite{kasica_table_2021}. All three operations suggest an interaction on two tables that results in one table. Thereby, we found one gesture (i.e., \emph{merge}, see Fig.~\ref{fig:teaser}) can meet the semantic requirements of all three operations.
The same applies to \emph{subset} and \emph{decompose}, and we used a combination of \emph{filter+extract} gestures for them.
Deleting tables, rows, and columns shares a similar gesture as the metaphor of throwing things away but differs in the target selection.

\begin{figure*}
    \centering
    \includegraphics[width=0.8\textwidth]{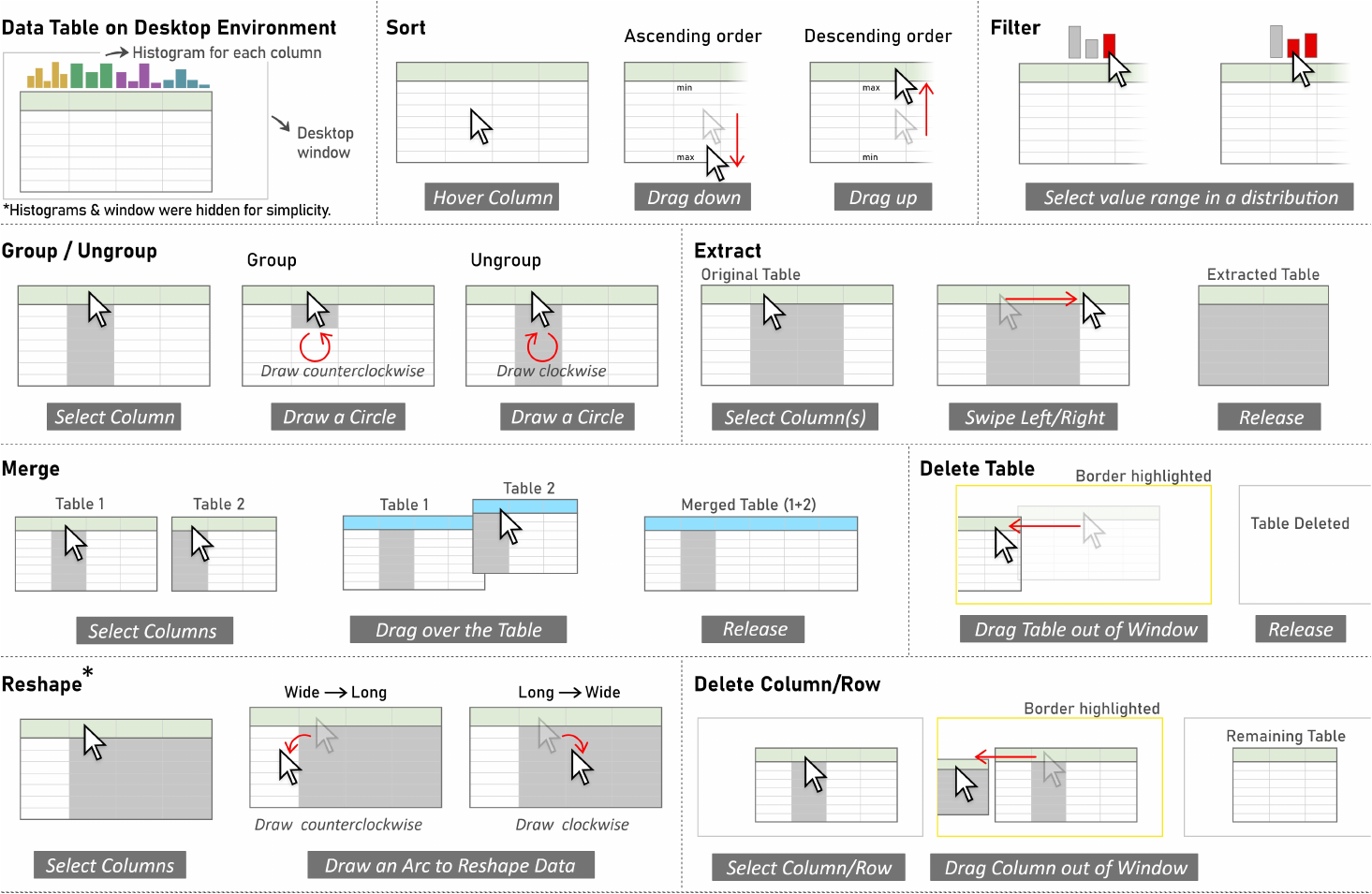}
    \vspace{-3mm}
    \caption{Desktop data table and gestures for data transformation tasks in the Desktop+Gesture condition. The operation that was excluded from the study is marked with *.} 
    \label{fig:desktopgesture}
    \vspace{-4mm}
\end{figure*}

We iterate the gesture designs among the team members. 
The design objective is to ensure the gestures can intuitively reflect their semantic meanings of data transformation and that no conflicts exist between different gestures.
\added{Some initial gesture designs were inspired by the teaching materials of an undergraduate data science course taught by two co-authors, where they frequently used gestures as metaphors for data transformation operations.
Specifically, within a data table, horizontal movements were naturally linked to column operations, while vertical movements were considered row operations. Meanwhile, operations involving multiple data tables were intrinsically demonstrated as two-handed gestures.
This initial design covered a good range of operations summarized by Kasica et al.~\cite{kasica_table_2021}. 
We further introduced more gestures to increase the coverage of data transformation operations, Table~\ref{table:operations}.
}
In summary, we have designed the following gestures:

\vspace{-1em}
\begin{itemize}[leftmargin=*]
  \item \textbf{Extract}: after selecting the target row(s) or column(s), \emph{pull} them out from the original data table to create a new table with the selected content. The original table will be kept, and a new table will be created.
  
  \item \textbf{Merge}: \emph{move} one table to \emph{collide} with another table and \emph{release} to combine two tables into one. If the key columns~\cite{kandel_wrangler_2011} are selected and match the criteria, an operation similar to \emph{JOIN} in SQL will be performed. If no key columns are selected, and two tables share the same structure, an operation similar to \emph{UNION} in SQL will be performed. Otherwise, the two tables cannot be combined. After merging, the original tables will be kept, and a new merged table will be created.
  
  \item \textbf{Filter}: for each column, a histogram is created to show its distribution. \emph{Brush} the histograms to select the range of values to keep. The original table will be updated.
  
  \item \textbf{Sort}: \emph{swipe} at a column from top to bottom or bottom to top to rearrange data in ascending or descending order. The original table will be updated.
  
  \item \textbf{Group/Ungroup}: after selecting the target column(s), \emph{squeeze} to aggregate the values, while \emph{expand} to restore the aggregated values to their original values. The original table will be updated.

  \item \textbf{Reshape}: after selecting the target column(s), \emph{rotate clockwise} to transform LONG data shape to WIDE shape, while \emph{rotate counterclockwise} to transform WIDE data shape to LONG shape. 
  For LONG to WIDE, the values in target column(s) will be grouped into key-value sets, while for WIDE to LONG, the values in target column(s) will be categorized. The original table will be updated. 
  
  \item \textbf{Delete}: after selecting target row(s), column(s) or table(s), \emph{throw} them away to remove the selected content. The original table will be replaced.
\end{itemize}

Our gesture design on Desktop shares some similarities with the GestureDB system~\cite{nandi2013gestural,jiang2013gesturequery}. We further adapted and extended those gestures into embodied interactions in VR.
Most gestures require only one input device (i.e., for \emph{extract}, \emph{merge}, \emph{filter}, \emph{sort}, \emph{reshape}, and \emph{delete}). Those gestures are almost identical in interaction behaviors on Desktop and VR (i.e., consist of actions like click, drag, and drop).
In VR, the ability to use two input devices (i.e., left and right hand-held controllers, which are visually represented as hands in our prototype) provides an alternative way to \emph{merge} data tables: people can manipulate two data tables simultaneously by grabbing one in each hand and moving them close to \emph{merge} them. 
The \emph{group/ungroup} gestures also leverage the two input devices in VR to \emph{squeeze} and \emph{expand}, while we use the \emph{draw a circle} counterclockwise and clockwise to imitate the same semantic meaning on the Desktop due to only one input device is available.

The designed gestures for VR are illustrated in Fig.~\ref{fig:vrgesture}, and the desktop gestures are presented in Fig.~\ref{fig:desktopgesture}.
\added{We conducted a pilot study with three computer science graduate students who have data science experience to verify the usability of our designed gestures. Throughout this pilot, we confirmed the feasibility of our chosen gestures to complete data transformation tasks. The pilot study revealed no major operational issues and confirmed the intuitiveness of our gestures.}


\section{User Study}
\label{sec:study}

This study involves two primary experimental factors: the computing environment (or \fEnvironment{}, i.e., Desktop vs. VR) and the interaction method (or \fInteraction{}, i.e., WIMP vs. Gesture), see Fig.~\ref{fig:condition_matrix}. 

\subsection{Study Conditions}
\label{sec:conditions}

\begin{figure}
    \centering
    \includegraphics[width=0.7\columnwidth]{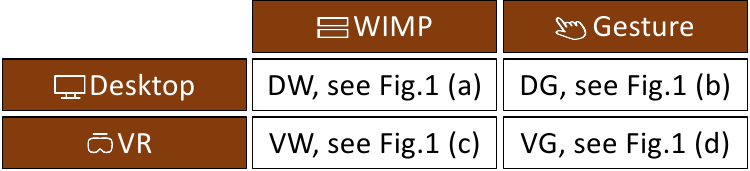}
    \vspace{-3mm}
    \caption{Our study compares two primary factors, leading to four conditions.}
    \label{fig:condition_matrix}
    \vspace{-5mm}
\end{figure}

To systematically investigate the two primary experimental factors, we included four conditions that cover all their interactions (Fig.~\ref{fig:condition_matrix}), namely, \desktopWIMP, \desktopGesture \vrWIMP, and \vrGesture (Fig.~\ref{fig:teaser}).
The conditions are also demonstrated in the supplemental video. 

\textbf{\fDesktopBold{} Conditions}:
On the Desktop, we provide an \emph{infinite canvas} as the primary working space. The infinite canvas is a \emph{zoomable} and \emph{pannable} canvas with no boundary where the user can place and move digital content (data tables, in our case). 
It provides extra freedom to content organization and overcomes the size limitation of a physical screen.
Various commercial tools (e.g., Miro~\cite{miro}, Google Jamboard~\cite{jamboard}, Microsoft Whiteboard~\cite{whiteboard}, and SAGE~\cite{marrinan2014sage2}) use it as their working space, and there are also a series of attempts of using it in data science~\cite{kodagoda_using_2013,elmqvist_datameadow_2007,javed_explates_2013,marrinan2014sage2}.
Following their success and design, we allow the user to zoom in and out of the workspace by scrolling the mouse scroll wheel. 
The same interaction has been implemented in many widely used Zoomable User Interfaces (ZUI), like Google Maps.
The user can also move data tables to any desired location using drag\&drop. 
A mouse is the input device for the two Desktop conditions.
In \emph{Desktop+WIMP}, the user clicks buttons to trigger operations (Fig.~\ref{fig:teaser}(a)). 
We provide one button for each operation and place all buttons on a panel that is fixed on the right side of the screen. It is visible to the user all the time.
In \emph{Desktop+Gesture}, the user uses our implemented gestures (Fig.~\ref{fig:desktopgesture}) to perform operations (Fig.~\ref{fig:teaser}(b)).

\textbf{\fVRBold{} Conditions}:
In VR, we allow the user to physically move in the space and freely place and move data tables to any location around them.
In \emph{VR+WIMP}, operation buttons were placed on a panel and interacted in the same way as in Desktop+WIMP. 
Due to the different display spaces between Desktop and VR, it is unclear where to place this panel.
To make the panel placement in VR as close as the Desktop+WIMP condition, we initially placed it using a head-reference approach~\cite{lages2019walking}, i.e., the panel will move as the user rotate their head, always visible to the user.
However, our preliminary test indicates that such a design is distracting and annoying.
We then changed the design to attach the panel to a left-hand-held controller according to~\cite{yang2020embodied}, where the user can easily access or hide it with arm movements.
The latter design was clearly preferred by the users we tested with, and was used in the user study (Fig.~\ref{fig:teaser}(c)).
We also noticed participants struggled to select rows and columns precisely in our pilot study.
Participants' comments revealed the need for a real-time visual indicator for selections. 
Thus, we rendered a ``red dot'' to indicate the pointer position on the data table. 
This mitigated the difficulties in selecting rows and columns based on another round of pilot tests.
In \emph{VR+Gesture}, the user uses our implemented gestures (Fig.~\ref{fig:vrgesture}) to perform operations (Fig.~\ref{fig:teaser}(d)).


\textbf{Summary}:
WIMP and Gesture differ in the way they trigger the operations. Additionally, performing operations in WIMP requires the user to move the cursor or pointer back and forth between the data tables and menu panel, which is likely to introduce a context-switching cost, while the gestures are directly operated on the data tables.
Desktop and VR both have an ``infinite'' display space and let the user reposition the table at any location.
Regarding the navigation method, the Desktop provides pan\&zoom, while VR enables physical navigation.

\subsection{Participants}
We recruited 20 participants (Male=16, Female=4; Age from 18 to 35) from the university mailing list after screening for their data transformation experiences with a five-question quiz. 20 out of 25 respondents answered at least four questions correctly and were invited to participate in the study.   
Nine participants indicated they use VR regularly on a weekly basis, another nine only used VR occasionally, and the rest two had no VR experience.
All participants had normal or corrected-to-normal vision.
We provided a \$20 Amazon Gift Card as compensation for each participant.

\subsection{Experimental Setup}
For VR conditions, we used a Meta Quest 2 virtual reality headset with $1920\times1832$ resolution per eye and a 90~Hz refresh rate. 
\added{For Desktop conditions, we used a 27'' monitor with a $2560\times1440$ resolution and 75~Hz refresh rate, which is a standard office setup.}
Both conditions use a PC with an Intel i7-11800H 2.30 GHz processor and NVIDIA GeForce RTX 3070 graphics card. 
We used the Air Link 
feature from Meta, which uses the PC for computation and the headset for rendering.
Air Link enables a wireless experience while still leveraging the stronger computing power from the PC.
Meanwhile, participants could move around more confidently without worrying about being tripped by cables.
The study took place in the space of 3.5 x 3.5 meters ($12.25m^2$), and \added{we let participants freely walk around the given area in VR conditions.
The participants were asked to place themselves in the center of the actual space at the beginning of every VR condition.}
In Desktop conditions, participants sat on a comfortable office chair with the monitor placed in front of them on an office desk.

On the Desktop, every initial data table, including the target data table, was the same size as $530\times400$ pixels. The initial tables form a grid layout to maximize the use of display space, and the target table was placed in the middle for participants to reference and remember easily.
In VR, five initial data tables, each with the size of $1.15\times0.65$ meters, were placed $1.65$ meters in front of the participant's initial position in a \added{semi-circular curved layout, which was identified as an effective space use strategy in VR for multi-window applications~\cite{liu2020design,satriadi2020maps}, which allows all data tables to be placed at the same distance within participant's reach.
The target table in VR was placed $35$ cm higher than the initial table for the same purpose as in the Desktop conditions, i.e., to be easily distinguishable from other tables. 
The initial data tables and the target data table cannot be deleted. This is to avoid accidentally deleting those data tables and being unable to complete the task.
}

\subsection{Task and Data}
\label{sec:task_and_data}
We asked our participants to \emph{perform data transformations with five given data tables to produce a table in the target format}.
\added{Participants could move and resize all the given data tables and the target data table.} 
To decide the number of given data tables, we piloted three different options (i.e., three, five, and eight).
We found the task was too obvious with three tables and too difficult with eight tables, so we decided to use five tables for the study.

After training, the participants were asked to use data transformation operations to complete the task with each condition. 
Initially, we included all operations from Sec.~\ref{sec:operations}. 
However, in our pilot tests, we found participants had a hard time conceptually understanding and applying the \emph{reshape} operation, resulting in an unexpectedly long completion time for each trial ($>$30 minutes).
This was aligned with some previous works~\cite{kandel_wrangler_2011,kasica_table_2021,raman2001potter}, pointing out the \emph{reshape} operation can be too complicated for novice users.
Thus, we decided to remove the \emph{reshape} operation from the task to target non-technical data workers and control the study duration.
We further conducted another test without using \emph{reshape}. 
Participants could complete the task in a reasonable amount of time (around 15 minutes per trial) without struggling.
To ensure the difficulty of each trial was similar, all trials required a minimum of twelve operations. The sequence of the required minimum operations in each trial was different to reduce the learning effect.

We used tabular data sets collected online for training and study tasks. 
\added{To eliminate the effect of participants' previous experience, we explicitly told them the data was not reflective of real-world information.}
We also controlled the data size of initial data tables (row count for each: 30, total column count: $\sim$20) and target table (row count: 10, column count: $\sim$6). 
We have included all our study stimuli in the supplementary material.




\subsection{Design and Procedures}
The conducted user study followed a full-factorial within-subjects design. 
We used a Latin square (4 groups) to balance the study conditions. 
Each participant completed four study trials, i.e., one for each condition.
The user study lasted two hours on average. 
The participants were first welcomed and reviewed to sign a consent form.
Participants were then instructed about the purpose and steps of the study. They will then complete the following components of the study:

\textbf{Adjustment}:
\added{We asked participants to adjust the Quest 2 headset (e.g., the IPD) to a comfortable setting for the VR conditions. Similarly, for the Desktop conditions, participants were instructed to adjust the chair height to their preference before starting the tasks.}
We confirmed that all participants could see the sample text clearly in all conditions before proceeding. 

\textbf{Training}:
We first introduce the data transformation terms, considering people might use different terminologies for the same data transformation.
We then introduced the computing environment (i.e., Desktop or VR) when a participant first encountered it. 
Sufficient time was provided for them to get familiar with the hardware until the participant asked to continue (usually around five minutes).
For each study condition, when it first appeared to the participant, we first asked them to watch a video demonstrating each operation in that study condition.
We confirmed that the participant fully understood how to perform each operation before moving to the next one.
After that, we asked participants to perform the same task as in the user study but with only three initial data tables and one target data table.
In this phase, we encouraged participants to ask  questions about interactions and study tasks. 
All participants completed the training by finishing the task and confirmed familiarity with the study conditions (the training task took around five to seven minutes). 



\textbf{Study Task}:
The study task with each study condition started after participants completed the training session for that condition. 
\added{Before we started the study, we ensured participants were well-informed by providing sufficient context. This included a brief overview of the datasets and the high-level semantic meaning of the target data table.
Participants had no time limit for task completion, but we instructed participants to complete the task as accurately and as fast as they could.}
For the VR environment, we repositioned participants to the room's center and let them face the same direction before each study task. All participants were able to complete the study task.


\textbf{Questionnaires}.
\emph{Post-condition questionnaires:} after completing the study task with each condition, participants were first asked to recall their performed operations in sequence. They were informed about this question at the beginning of the user study.
They then filled out a Likert-scale survey adapted from SUS and NASA TLX to rate their subjective experience and provide
qualitative feedback about the pros and cons of that condition.
\emph{Post-study questionnaires:}  after completing all study tasks and post-condition questionnaires, participants were asked to rank all study conditions based on their overall experience. We asked them to provide demographic information at the end.


\subsection{Measures}

We collected quantitative data and interaction records for each study condition to capture their task performance and sense-making process. 
Specifically, we used the following measures.
\added{\textbf{Error score}: we compared the difference between the target table and the participant's result table by rows and columns. Each difference contributed to one error score. The order of rows was considered (as \emph{sorting} operation was included in the task), while the column order did not affect the error score.}
\textbf{Time}: we measured the time from the initial data tables that were first rendered to the participant's task completion.
\textbf{Number of operations}: we recorded the total number of operations performed by participants to complete the task.
\textbf{Recall score}: we calculated the \emph{Levenshtein distance} between the participant's actual performed operations sequence and recalled operations sequence that were collected in the questionnaire. A lower value means a closer match and suggests participants can remember their action history more accurately.
\textbf{Number of performed delete operations}: we recorded the total number of performed table deleting operations in each study trial.
\textbf{Number of data tables left:} we documented the number of data tables when the participant completed each study trial.

We also collected subjective ratings on a seven-point Likert scale for \textbf{mental demand}, \textbf{physical demand}, \textbf{learnability}, \textbf{engagement}, and \textbf{usability}. Lower mental and physical demands were considered positive, while higher learnability, engagement, and usability were treated as beneficial. Participants also \textbf{ranked} their overall experience.
\textbf{Qualitative feedback} about each condition's pros and cons were collected from participants. Two authors derived a set of codes from the responses of the first five participants and applied the codes to the remaining responses. 


\subsection{Hypotheses} 
\label{sec:hypo}
We developed our hypotheses based on previous empirical results and our analysis of study conditions, see Sec.~\ref{sec:conditions}. 

\textbf{Error score}. We did not expect any difference in the error score as all required operations were provided consistently across all conditions. We believed that participants could complete the study task for all conditions successfully.

\textbf{Time}. We expected Desktop to outperform VR ($H_{time-env}$) based on previous studies, which found desktop interactions faster than VR interactions due to less required movement~\cite{bach2017hologram,chen2012effects,wagner2018immersive,arms1999benefits}.
Meanwhile, we anticipated Gesture to be faster than WIMP ($H_{time-interaction}$) considering better completion time of embedded/embodied interaction over WIMP found in earlier research~\cite{drucker_touchviz_2013,sarvghad_embedded_2019}. 

\textbf{Number of operations}. 
Our tasks require a sense-making process of foraging and structuring information to solve the problem.
Under such a context, we considered VR requires a fewer number of operations than Desktop ($H_{ops-env}$) based on the previous investigation of sense-making in immersive space~\cite{lisle_evaluating_2020,lisle_sensemaking_2021}. 
We also expected Gesture requires a fewer number of operations than WIMP ($H_{ops-interaction}$). With a lower context-switching cost, Gesture would have fewer disruptions from navigation~\cite{ball2005analysis} and require less ``mental map'' rebuilding for the users~\cite{purchase2006important,reda2015effects}.

\textbf{Recall score}.
We expected participants to have a better recall performance in VR than Desktop ($H_{recall-env}$), as VR with a 3D spatial environment was found to be more effective than Desktop for memorizing 
and retrieving information~\cite{yang2020virtual}.
Meanwhile, we foresaw Gesture outperforms WIMP ($H_{recall-interaction}$) since performing body motions with the Gesture has a positive effect on memorability~\cite{zagermann2017memory}.

\textbf{Number of performed delete operations} and \textbf{data tables left}.
We predicted a fewer number of delete operations ($H_{del-env}$) and a larger number of data tables left ($H_{left-env}$) in VR than in Desktop. 
We argue that the larger display space in VR allows participants to keep more intermediate results and reduce the need to delete them.
We also anticipated Gesture requires a fewer number of delete operations ($H_{del-interaction}$) and has a larger number of data tables left ($H_{left-interaction}$) than WIMP. 
As discussed, Gesture has a lower context-switching cost than WIMP, which can reduce the content organization workload and increase the maximum number of data tables participants can handle.

\section{Results}
\label{sec:results}

We present our statistical results regarding our hypotheses, outline participants' strategies for using the display space and summarize qualitative feedback for each condition.
For dependent variables or their transformed values that met the normality assumption, we used \emph{linear mixed modeling} to evaluate the effect of independent variables on the dependent variables~\cite{Bates2015}. 
Compared to repeated measure ANOVA, linear mixed modeling does not have the constraint of sphericity~\cite[Ch.\ 13]{field2012discovering}.
We modeled all independent variables (\fEnvironment{} and \fInteraction{}), and their interactions as fixed effects. A within-subject design with random intercepts was used for all models. 
We evaluated the significance of the inclusion of an independent variable or interaction terms using a log-likelihood ratio. 
We then performed Tukey's HSD posthoc tests for pairwise comparisons using the least square means~\cite{Lenth2016}. 
We used predicted vs. residual and Q---Q plots to graphically evaluate the homoscedasticity and normality of the Pearson residuals respectively. 
For other dependent variables that cannot meet the normality assumption, we used the \emph{Friedman} test to evaluate the effect of the independent variable, as well as a Wilcoxon-Nemenyi-McDonald-Thompson test for pairwise comparisons. Significance values are reported for $p < .05 (*)$, $p < .01 (**)$, and $p < .001 (***)$.
We also report mean values, 95\% confidence intervals (CI), as well as Cohen's d as an effect size indicator for significant comparisons.

\begin{figure}
    \centering
    \includegraphics[width=1\columnwidth]{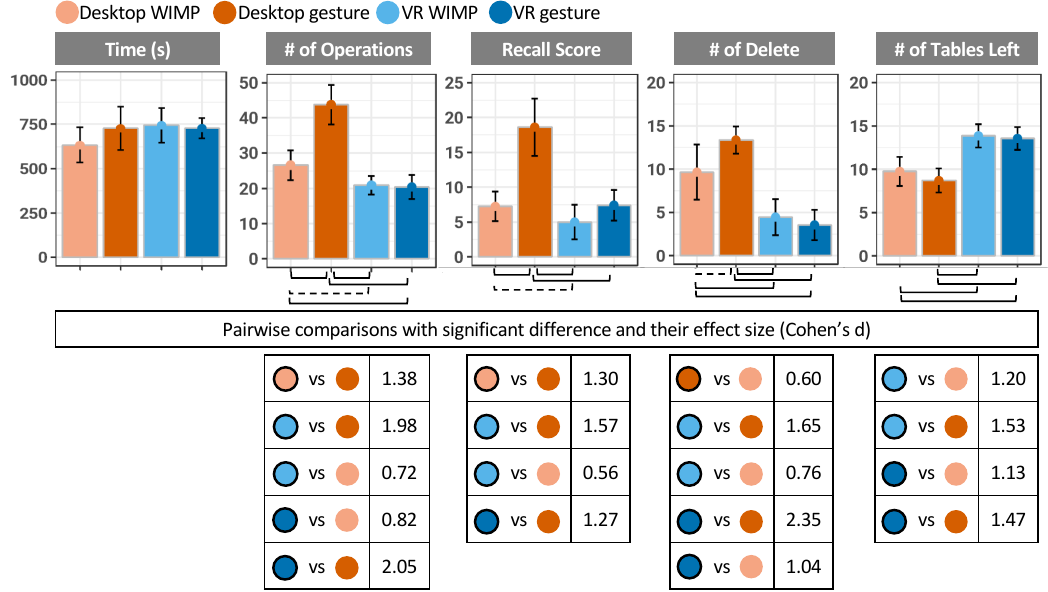}
    \vspace{-7mm}
    \caption{Measurement of time, the total number of operations, recall score, the number of delete operations, and the number of tables left by task. Solid lines indicate statistical significance with $p < 0.05$, and dashed lines indicate $p < 0.1$. The tables below show the effect sizes for pairwise comparison. Circles with black borders indicate the winning conditions.}
    \label{fig:performance}
    \vspace{-3.5mm}
\end{figure}

\subsection{Quantitative Results}
\label{sec:results-quan}
Results are illustrated in Fig.~\ref{fig:performance}, Fig.~\ref{fig:ratings}, and Fig.~\ref{fig:rank}.
All statistical analyses and anonymized data were included in the supplementary material.

\textbf{Error score.} As expected, all participants could complete the study task correctly under all conditions.

\textbf{Time.} Surprisingly, we found \fEnvironment{} ($p=0.14$), \fInteraction{} ($p=0.51$) and their interaction ($p=0.42$) did not have a significant effect on time. 
All conditions took similar amount of time: Desktop+WIMP (634s, CI=73s), Desktop+Gesture (727s, CI=134s), VR+WIMP (744s, CI=116s), and VR+Gesture (728s, CI=96s). Desktop+WIMP tended to be slightly faster but without statistical significance.
Thus, we reject $H_{time-env}$ and $H_{time-interaction}$.

\begin{figure}
    \centering
    \includegraphics[width=1\columnwidth]{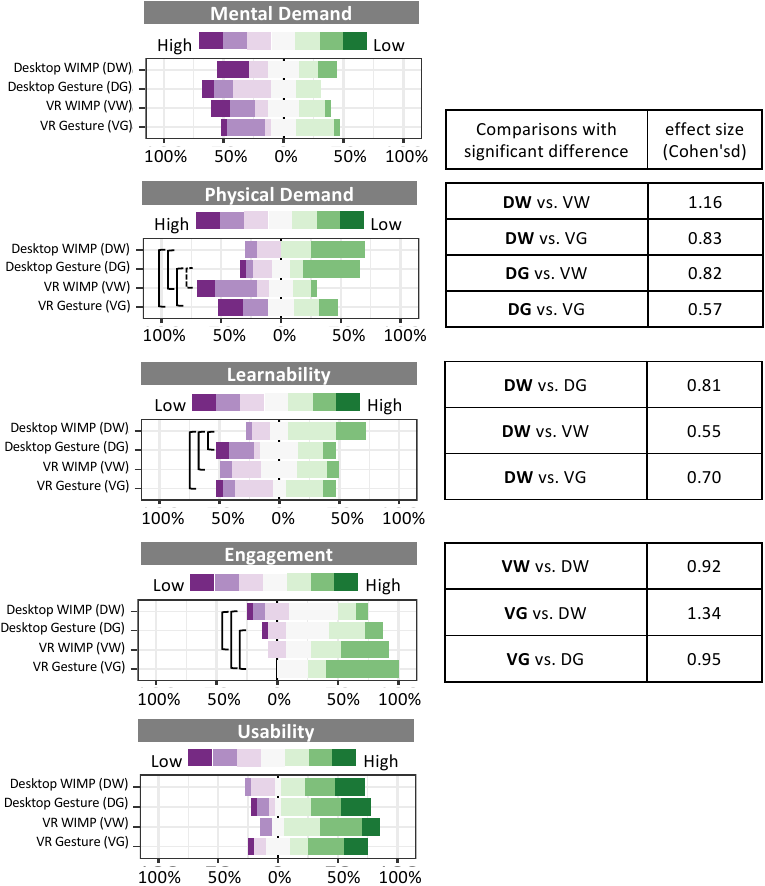}
    \vspace{-7mm}
    \caption{Subjective ratings on mental demand, physical demand, learnability, engagement, and usability by task. Solid lines indicate statistical significance with $p < 0.05$, and dashed lines indicate $p < 0.1$.}
    \vspace{-4.5mm}
    \label{fig:ratings}
\end{figure}

\textbf{Number of operations.}
We found significant effects of \fEnvironment{} ($***$), \fInteraction{} ($***$) and their interaction ($***$) on the total number of performed operations.
VR conditions (WIMP with 21.0, CI=2.90 and Gesture with 20.5, CI=2.54) required less number of operations to complete the study task than Desktop conditions (WIMP with 26.6, CI=4.22 and Gesture with 43.8, CI=7.06). All comparisons were statistically significant ($*$) except for the comparison between VR+WIMP and Desktop+WIMP ($p=0.057$).
On Desktop, Gesture also required more operations than WIMP ($***$).
In summary, we accept $H_{ops-env}$, and reject $H_{ops-interaction}$.

\textbf{Recall score.}
We found significant effects of \fEnvironment{} ($***$), \fInteraction{} ($***$) on the recall score, with a marginally significant effect from their interaction ($p=0.081$).
Condition-wise, participants can remember their action history better in VR+WIMP (5.0, CI=1.83), VR+Gesture (7.4, CI=2.20), and Desktop+WIMP (7.3, CI=1.94) than Desktop+Gesture (18.6, CI=5.42). VR+WIMP was also marginally more memorable than Desktop+WIMP ($p=0.078$).
For both the WIMP and Gesture, VR could better support the recall process than a Desktop.
Dekstop+WIMP was also found more memorable than Desktop+Gesture ($***$).
In conclusion, we accept $H_{recall-env}$, and reject $H_{recall-interaction}$.

\textbf{Number of performed delete operations} and \textbf{number of data tables left}. For the number of performed deleting operations and the number of left tables, there was a significant effect from \fEnvironment{} (all $***$). \fInteraction{} only had a significant effect on the number of performed deleting operations ($*$).
Participants deleted fewer tables and kept more data tables in VR than in Desktop (all $***$).
In summary, along with accepting $H_{del-env}$ and $H_{left-env}$, we reject $H_{del-interaction}$ and $H_{left-interaction}$.

\textbf{Ratings.}
We found a significant effect of the study condition on \emph{physical demand} ($***$), \emph{learnability} ($**$), and \emph{engagement} ($***$).
Participants found VR conditions (WIMP with 4.0, CI=0.71 and Gesture with 3.7, CI=0.92) more physically demanding than Desktop conditions (WIMP with 2.25, CI=0.69 and Gesture with 2.6, CI=0.88). 
Participants considered Desktop+WIMP (4.7, CI=0.55) most easy to learn, with Desktop+Gesture (3.45, CI=0.81), VR+WIMP (4.0, CI=0.55), and VR+Gesture (3.7, CI=0.76).
VR+Gesture (6.4, CI=0.41) was found more engaging to use than Desktop conditions (WIMP with 4.7, CI=0.73 and Gesture with 5.3, CI=0.64).

\textbf{Ranking.} 
Participants ranked the VR+Gesture as providing the best overall experience, with 65\% ranked it as first place, and 25\% ranked it as second place (i.e., 90\% in total ranked VR+Gesture as the first or second place).

\subsection{Layout Strategies}

\begin{figure}
    \centering
    \includegraphics[width=0.7\columnwidth]{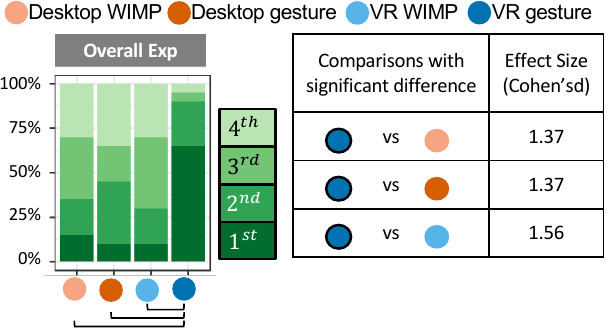}
    \vspace{-3mm}
    \caption{User ranking of overall user experience for each condition. Solid lines indicate significant differences with $p < 0.05$. 
    }
    \vspace{-4.5mm}
    \label{fig:rank}
\end{figure}

To better understand how participants used the display space, we grouped the final layout of each trial.
We found participants had different strategies in Desktop and VR, but used a similar layout between Gesture and WIMP.
All final layouts are included in the supplementary material.

\begin{figure}
    \centering
    \includegraphics[width=1\columnwidth]{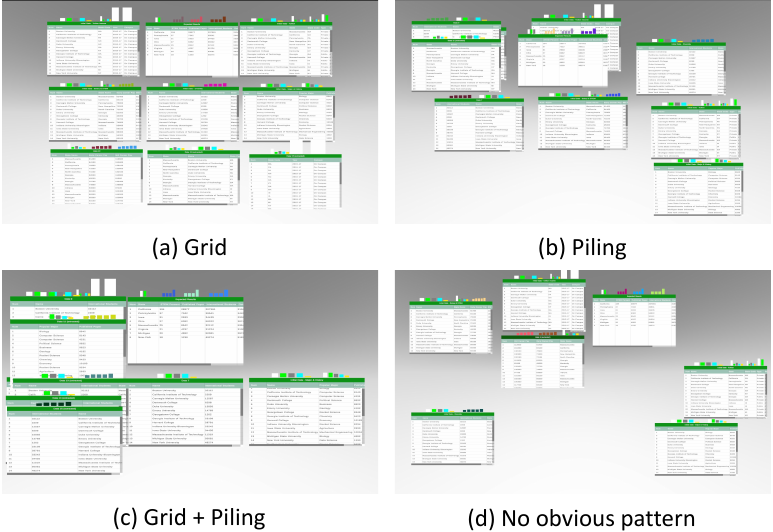}
    \vspace{-7mm}
    \caption{Four layout strategies used by our participants on \fDesktop{}.}
    \label{fig:organize_desktop}
\end{figure}

On Desktop, we identified four different layout strategies (Fig.~\ref{fig:organize_desktop}):
\textbf{Grid} (five). Participants placed their data tables in a regular layout, closely forming a grid shape.
\textbf{Piling} (one). The participant created a few piles of data tables.
\textbf{Grid+Piling} (five). The participants created a regular layout with some piles of data tables.
\textbf{No obvious pattern} (nine). Roughly half of the participants did not demonstrate a clear layout pattern and created an ``organic'' layout.







\begin{figure}
    \centering
    \includegraphics[width=1\columnwidth]{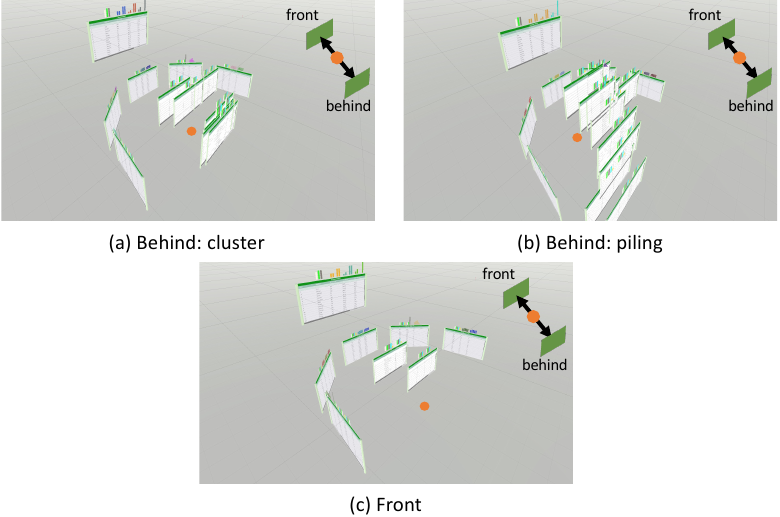}
    \vspace{-7mm}
    \caption{\added{Three layout strategies used by our participants in \fVR{}. The red dot indicates the position of the participant.}}
    \vspace{-4.5mm}
    \label{fig:organize_vr}
\end{figure}

In VR, 19 out of 20 participants almost did not move the five initial tables or the target table. Participants seemed to treat these given tables in 
as \emph{strong anchors} and were reluctant to manipulate them. 
This observation might reflect participants' intention to keep data provenance. 
Thus, VR might increase the awareness of provenance and allow the track of provenance to be more manageable.
Regarding the final layouts, we found three different strategies (Fig.~\ref{fig:organize_vr}).
17 participants performed data transformation \added{behind themselves:} 
\textbf{Behind-cluster} (16). Participants created a few clusters at the back of their initial orientation.
\textbf{Behind-piling} (one). Participants piled data tables in one cluster and formed a roughly vertical line at the back of their initial orientation.
\textbf{Front} (three). Participants used the space in front of their initial orientation, and to avoid occlusions, they had to delete data tables more frequently.
In all strategies, we observed that participants preferred to stay in the center  and place data tables close to them to reduce physical movements.

Due to limited data points in some strategies, we could not identify significant performance differences between different strategies, except that the \emph{Front} strategy required more \emph{deletion and total operations} and had fewer \emph{tables left} than other strategies utilizing the entire 360\textdegree{} circular space in VR.

\subsection{Qualitative Feedback}
We performed qualitative coding to extract common themes from user feedback on each condition. For each condition, we listed the top three mentioned codes and those mentioned more than five times by the participants (frequency shown in parenthesis). We further highlighted the top codes with other frequently associated codes.
Finally, we summarized the overall insights across all conditions.
The complete coding results can be found in the supplemental materials.

\textbf{\desktopWIMPBold{}} was considered \textit{straightforward} (7), and \textit{familiar} (5). The downsides were \textit{limited space} (11), \textit{hard to interact} (6), and \added{button interaction were} \textit{not intuitive} (5). Specifically, limited space was the primary concern, as shown by its association with several codes, including hard-to-interact (3) and 
\added{the constant shift in focus from the working window to the buttons created task disruptions when executing operations (2).} 

\textbf{\desktopGestureBold{}} was considered \textit{intuitive (15)}, \textit{better than button} (8), and \textit{easy to use} (8). Because of the intuitive feeling, it was considered better than buttons and easy to use by five users, good for merging by four users, and good for sorting by two users.  
\added{The primary issues included \textit{limited space} (9), \textit{hard to interact} (8), \textit{discoverability issues} (6), and \textit{functionality} (5). Similar to the Desktop+WIMP, the limited space was the predominant concern linked to other codes. This included instances where gestures were occasionally challenging to execute due to the limited space, which resulted in creating task disruptions (3), particularly when executing certain functions (3) that demanded more space, such as extraction tasks.}

\textbf{\vrWIMPBold{}} was praised with \textit{large space} (12), \textit{better than desktop} (9), \added{\textit{grabbed\&moved} navigation was} \textit{intuitive} (7), \textit{easy to use} (6), and \textit{flexible} (6). Specifically, the major benefit of the large space was associated with \added{flexibility} by five users, easy to use by two users, and good understanding by two users. Among comments on better than desktop, people referred to it as easy to use (5), easy to organize (2), and accessible (2).
\added{Conversely, the issues related to \textit{functionality} (13).
In addition, one participant pointed out that using a pointer to select two tables from a dropdown menu caused greater disruption when trying to execute merge operations compared to the Desktop+WIMP setup. The concerns arose regarding the constant need to hold buttons on the left hand, causing task disruptions (6), along with another reason similar to those encountered with the Desktop+WIMP. 
People had diverse opinions on the functionality, such as hard-to-select (3), heavy headset (2), and other technical issues like resolutions.}

\textbf{\vrGestureBold{}} was found to be \textit{intuitive} (17), \textit{better than button} (10), \textit{easy to use} (7), \textit{straightforward} (5), and \textit{flexible} (5). The majority (17) found it intuitive and subsequently associated it with better than buttons (7), easy to use (6), and flexible (4). Three also cited more accessible, and the other three cited large spaces. Furthermore, people considered it better than buttons mainly because it is easy to use (4), straightforward (4), accessible (3), and promotes good understanding (2). 
\added{Participants particularly spoke highly about gesture interactions and grab\&move navigation.}
On the other hand, the major concern was \textit{functionality} (12), mostly with resolution and technical issues (7), \added{but no specific functionality issues were found to perform operations in VR+WIMP.}

\textbf{Summary.} All conditions were considered \textit{easy to use}. In Gesture conditions (Desktop+Gesture and VR+Gesture), the majority cited \textit{intuitive} (15 and 17 times, respectively) and considered them \textit{better than button} (8 and 10 times). 
VR conditions (VR+WIMP and VR+Gesture) were considered \textit{flexible} (6 and 5 times) and \textit{better than desktop} by several users (9 and 4 times).
Interestingly, Desktop+WIMP and VR+Gesture each were cited \textit{straightforward} by a handful of users (7 and 5 times), indicating that users have expected that WIMP is natural for desktop and 3D gestures for VR environment. More discussions were presented in Sec.~\ref{sec:results}.

On the other hand, Desktop conditions 
hinder interaction due to \textit{limited space} (11 and 9 times) and \textit{hard to interact} (6 and 8 times).
With VR conditions, 
the majority have commented on \textit{functionality} (13 and 12 times) due to the unfamiliarity with data transformation in VR, such as technical issues and feature suggestions. In VR+WIMP, people found it \textit{hard to interact} and \textit{select} (5 times), leading to creating \textit{disturbed} (6 times).


\section{Key Findings and Discussions}
\label{sec:discussion}


\textbf{Performing data transformation in \fVRBold{} and on a \fDesktopBold{} had similar time performance.}
Previous studies produced mixed results in comparing VR and Desktop. 
In immersive analytics, for completion time, VR was found to be primarily beneficial for visualizing spatial data and 3D shape perception~\cite{fonnet2019survey,ens2021grand}, and to be slower than Desktop due to more required movements (see our \emph{time} hypothesis in Sec.~\ref{sec:hypo}).
Our tested task did not involve perceiving 3D or spatial visualization, so we expected Desktop to be faster than VR. Surprisingly, our four tested conditions had similar time performance.
We believe the empirical results of VR being slower than Desktop in performing interactions still applied to our case, which is partially reflected by the fact that VR conditions were considered significantly more physically demanding than Desktop conditions.
On the other hand, unlike the previously tested tasks, our tested task not only involved low-level interactions but also required participants to actively make sense of the data to come up with a sequence of interactions to complete the task. 
We could reasonably expect participants to put a significant amount of effort into thinking and planning strategies for a complicated task like ours.
Based on this assumption and our results, we suggest that VR allows participants to complete the high-level sense-making components faster than Desktop, which supplements its extra time costs in performing low-level interactions.
Below, we elaborate more on the high-level sense-making process from the \emph{provenance} and \emph{strategic thinking} perspectives.


\added{\textbf{\fVRBold{} 
showed the potential to provide improved provenance over \fDesktopBold{}}.
Provenance is about the lineage and processing history of data. 
\added{It provides a detailed record of the origins of the data, how it has been processed or modified, and where and when it transformed over time~\cite{wang2015big}.}
The ability to track provenance can support many applications, like data quality control, data auditing, and replication~\cite{simmhan2005survey}.
In our study, we used the \emph{recall score}, \emph{the number of performed delete operations}, and \emph{the number of data tables left} as indicators of certain aspects of provenance. The \emph{recall score}  measures the ability to recall the operation history, reflecting when the data was transformed. The latter two metrics offer objective measures of the amount of kept information.}

In addition to the previously confirmed memorability advantage in VR (see our provenance hypothesis in Sec.~\ref{sec:hypo}), we believe the large display space and embodied navigation in VR also helped participants keep track of provenance.
\emph{First}, participants had more space to place tables in VR, whereas on Desktop, they were more likely to delete tables to free display space to reduce clutter. 
Our results of fewer table deletions and more kept tables in VR than on Desktop are well aligned with provenance tracking ability.  
\added{The data tables, kept persistently visible, provided a reliable reference for participants to confirm the success of their current operations. Moreover, they were a handy tool for participants to return to whenever they made mistakes. This was likely due to the large display space provided within the VR setup.}
\added{Particularly, nine participants explicitly complained about the display space on the Desktop; for example, \textit{``I have a very limited workspace (on the desktop), and cannot see all the tables at once, which really hurts my performance (P15).''}}
\added{This limited space on the Desktop lets participants continually delete the data tables, which leads to a loss of opportunities to keep track of the processing history.}
\emph{Second}, we consider physical navigation more efficient than virtual navigation.
Previous research identified the benefits of using physical movements to navigate large displays over virtual navigation (i.e., zooming)~\cite{ball2005effects,ball2007move}.
Our results partially re-confirmed their findings in VR.
12 participants' comments reconciled with our assumption, like,  
\textit{``I have all data tables in front of me without zooming in \& out, and I can focus on the task (P12).''}
\added{We found that the participants prefer physical navigation, as it offers a more intuitive and faster method for altering the perspective view of their workspace for accessing all kept information.}
One participant also specifically described the use of large display space to improve spatial memory and embodied navigation in VR: 
\textit{``I like the point that I put the data table behind me so that I can come back if I make mistakes (P7).''}
\added{This feedback aligns well with the improved provenance in VR, which enhances data quality through auditing without manually documenting every detail of each step.}
\added{However, provenance is still a relatively abstract concept, and our measurements also only captured certain aspects of it. More effort is needed to formally define and quantify provenance in data transformation.}

\textbf{\fVRBold{} demonstrated preliminary evidence of promoting strategic thinking.}
Participants needed to continuously develop the next steps in our tested task and evaluate their progress.
We found VR required fewer operations to complete transformation tasks than Desktop, which points to the potential advantages of VR in supporting strategic planning.
With the improved provenance, it is likely that VR users can better track the progress and take more efficient steps. Another potential reason is the flexible arrangement of VR tables that support easier visual comparison.
We observed \emph{all 20} participants in VR \emph{grabbed\&moved} their working table under the target table for comparison. 
Meanwhile, participants rarely performed a similar interaction on the Desktop.
We believe such frequent comparisons enabled continuous progress evaluation and promoted strategic thinking in VR.
We anticipate the embodied table management (i.e., the grabbed\&moved) and large display space in VR provide a natural mechanism to support such an approach. 
In contrast, the need for repetitive zooming in/out on Desktop made participants reluctant to perform the same strategy.
Additionally, participants also commented on their experiences along this line, for instance, 
\textit{``I have many more places to put the data table (in VR). Organizing the table was way more manageable (P14).''}
\textit{``The interaction in VR was way better than the monitor version; the task became way more accessible, and organizing was easy (P3).''}

\textbf{\fWIMPBold{} tended to be more suitable for \fDesktopBold{} than \fGestureBold{}.}
Unlike previous studies~\cite{drucker_touchviz_2013,sarvghad_embedded_2019}, we did not find positive effects on the collected measures of using Gesture over WIMP on a Desktop.
We believe the task and gesture complexities could be the main reasons for our contradictory results.
We tested a more complicated high-level task (average completion time was 10+ minutes per trial) than the previous low-level tasks (average completion time was around one minute per trial).
Performing a single interaction using Gesture could outperform WIMP in our study, but the complexity of Gesture might introduce other overheads that decreased its performance significantly in our tested task.
The subjective ratings partially confirmed our assumption: Desktop+Gesture  was considered harder to learn than Desktop+WIMP.
Participants had to remember more gestures (eight) than in the previous studies (three in~\cite{sarvghad_embedded_2019} and five in~\cite{drucker_touchviz_2013}), which might introduce a high working memory and affect their performance, especially in a more complicated task.
The fact that participants struggled with recalling their action history in Desktop+Gesture resonated with our conclusion, with some representative comments like:
\textit{``The interaction was confusing (in Desktop+Gesture). I think I would like[ly] have more errors than other conditions.''}.

\textbf{No noticeable difference in time, efficiency, and provenance measures between \fWIMPBold{} and \fGestureBold{} in \fVRBold{}.}
Similar to Desktop, the identified benefits of body motion~\cite{zagermann2017memory} might also be affected by the task and gesture complexities in VR.
On the other hand, we did not find any significant difference between VR+WIMP and VR+Gesture, indicating the overhead introduced by learning and remembering Gesture in VR could be negligible due to its intuitiveness. 
Participants ranked VR+Gesture with the best overall experience, and some commented on the benefits of using Gesture in VR explicitly, like
\textit{``Embodied interaction helps me a lot in understanding the data story (P2)''} and
\textit{``Using both hands was very helpful in performing the task (P4).''}
More importantly, the embodied table management (i.e., the grabbed\&moved) and physical navigation were provided in all VR conditions.
Compared to interactions for triggering the operations, these two features might contribute a heavier weight in the final performance, making both VR conditions have similar performances.
\added{Furthermore, the perceived intuitiveness of the gestures and embodied interactions was reflected in the feedback we collected.}
\added{Comparably, none of the participants explicitly mentioned Desktop+WIMP to be intuitive.}

\textbf{Various layout strategies exist in both environments.}
The ``infinite'' space in both the desktop and VR conditions allowed participants to freely lay out content in their workspaces. 
We found two distinct dimensions in their strategies: \textit{space usage} and \textit{view management methods}.
For space usage, on Desktop, most participants only used space slightly larger than the initial setup (Fig.~\ref{fig:organize_desktop}) with 6 to 9 data tables and adjusted the view by zooming. 
On the other hand, the majority of participants in VR utilized the entire 360\textdegree{} circular space around them to fit more data tables (12 to 15)
(Fig.~\ref{fig:organize_vr}).
In VR, previous studies observed using semi-circular layouts were more frequent~\cite{satriadi2020maps} and beneficial over 360\textdegree{} layouts~\cite{liu2020design,liu2022effects} for performing low-level tasks (e.g., search and comparison). 
However, for high-level tasks, our observations aligned well with some other works, such as sensemaking~\cite{davidson2022exploring} and visual exploration~\cite{batch2019there}, where people preferred using the entire fully circular space.
We anticipated that placing content within the field of view is beneficial for time-constrained low-level tasks, while fully utilizing the display space is essential for the larger amount of content generated in high-level tasks.
In terms of view management methods, the layout on the Desktop was considerably more organized than in VR, showing that precise view placement is easier with a mouse than with a VR controller. However, user behaviors were similar across both Desktop and VR. 
Most participants intended to create occlusion-free clusters (e.g., a grid layout in Fig.~\ref{fig:organize_vr}b), yet some participants created piled clusters (e.g., a piling layout in Fig.~\ref{fig:organize_vr}a), which was also observed in a recent AR study~\cite{luo2022should}. 
We also noticed the use of space to record the operation history from some participants (e.g., moving older tables to the top), which has been identified on large 2D displays~\cite{andrews2010space}.

\section{Generalizations, Limitations and Future Work}
\label{sec:limitations}

\textbf{Applications.}
Our study provides empirical evidence of the benefits of using VR for data transformation.
We believe the large display space, spatial memory, and embodied navigation offered by VR are the primary factors for its improved performance.
Since these are general characteristics of VR, we expect our results can be generalized to similar tasks that require organizing a large number of entities in space for sense-making.
Some preliminary research explored the use of VR in such applications, like multiple view visualization~\cite{cordeil2017imaxes,lee2020shared}, multi-scale geographic navigation~\cite{satriadi2020maps}, and large-scale document analysis~\cite{lisle_evaluating_2020,lisle_sensemaking_2021,luo2022should}. 
However, as highlighted by Ens et al.~\cite{ens2021grand}, there is a lack of fundamental study for comparing immersive versus non-immersive platforms for analytics purposes.
Our study provides a preliminary assessment for data transformation under this context. Future studies may test other applications and delineate the benefits of immersive environments for a broader range of analytical tasks.

\textbf{Target users and functionality.}
We focused on investigating intuitive data transformation tools for non-technical data workers and intentionally excluded the programming requirements in our study. However, programming-like operations are essential for more experienced users and more complicated tasks. 
The Gesture has limitations in how much information and intention it can express.
To extend this work in this direction, future work needs to integrate a programming interface into the prototypes, like many UI-based commercial tools (e.g., Tableau and Excel) and research prototypes~\cite{10.14778/3229863.3240493,shang2021davos}.
To achieve this in VR, we need to consider the most appropriate text input method~\cite{dube2019text}.
Alternatively, we may also consider using natural language as a user interface with higher 
expressiveness~\cite{Shen2022TowardsNL,narechania2020nl4dv}.
On the other hand, we also see opportunities to develop new interaction techniques (e.g., embodied gestures) to lower the learning curve of complicated data transformation operations for non-technical data workers.

\textbf{Techniques.}
We believe our designed gestures were intuitive and natural to their represented data transformation operations.
However, the Gesture did not contribute to the participants' performance as we expected.
We believe Gesture can still outperform WIMP in low-level tasks when the user knows the precise operation to perform~\cite{drucker_touchviz_2013,rzeszotarski2014kinetica,nandi2013gestural}.
Specifically, Nandi et al. found gestures to have better performance and discoverability than WIMP for a single data transformation operation~\cite{nandi2013gestural}.
However, discoverability might become a more severe issue in high-level tasks where the user must continuously develop the next steps.
Although during the training phase, the participants were able to complete training tasks with a smaller dataset, 
it was still reasonable for participants to spend extra time and effort to recall the gestures in the longer study tasks.
Future work may look at further improving the discoverability and learnability of gestures~\cite{bau2008octopocus,cockburn2014supporting,nacenta2013memorability} or conduct a longitudinal study to reduce this effect.

Additionally, there are a few other directions to extending our work.
Although we improved the selection experience in VR, precisely selecting and manipulating objects can still be challenging. Along with improving mid-air interactions, one may consider other input devices, like using a mouse in VR~\cite{zhou2022depth,pavanatto2021we} and other tangible proxies~\cite{cordeil2020embodied,satriadi2022tangible}.
Besides, we consider visualizing data processing history could further improve provenance, like a data flow system~\cite{javed_explates_2013,wright2006sandbox}. 


\textbf{Computing environments.}
We tested Desktop and VR, as Desktop is the most widely used, and VR is emerging. Testing other computing environments with different input modalities could bring more insights into interaction designs, like a tablet~\cite{nandi2013gestural, drucker_touchviz_2013, burley2019arquery, rzeszotarski2014kinetica} and display wall~\cite{badam2016supporting}.
Tablets offer multitouch capabilities and are considered more natural for performing gestures than Desktop. However, tablets are usually limited in size, and users may suffer from ``fat finger'' issues in precise interactions.
A larger touch screen may alleviate the issue, but not easily accessible to many people.
Nevertheless, testing different touching devices is an exciting future direction.
Moreover, specific to our study, our gestures primarily only involve click and drag, which could be easily completed with a mouse, as demonstrated in previous work~\cite{sarvghad_embedded_2019,saket2019investigating}.
\added{Increasing the display space on the Desktop is likely to increase the performance (like using multiple monitors or a display wall), as one of the most notable benefits of VR is its large display space.}
The effect of display size has been previously explored~\cite{ronne2011sizing}.
Future work can follow the same methodology to study the effect of display size on the data transformation workspace. 
\added{Furthermore, we found that our participants might not be familiar with our zoomable and pannable interface on the Desktop.
However, there's an increasing trend of no-code data science tools designed with zoomable and pannable interfaces, as we discussed in Section~\ref{sec:conditions}. It gradually becomes an essential design for non-technical data workers.
Despite this, we also see other opportunities for managing multiple data tables, such as using tabs.}

\textbf{Scalability and ecological validity.}
We tested relatively small data in our study (see Sec.~\ref{sec:task_and_data}), as we want to control the study duration. 
Meanwhile, we had an interesting comment from one participant that they did not actively check the rows of data tables but focused on the columns more.
This comment aligns well with the design of the GestureDB system~\cite{jiang2013gesturequery,nandi2013gestural}, where they present the columns all the time and only present rows on demand or for confirmation purposes.
Larger data should be tested in future studies, especially data with more columns.
\added{Meanwhile, it is challenging to display all columns for large data sets (e.g., with 20 columns). As such, in our tested data, we intentionally have columns that cannot fit into the table view. The width of each column was determined by its longest data entry, causing the total width of all columns to exceed the table's width. Hence, even with a limited number of columns, columns were not visible at once, and scrolling was necessary to finish the task.}
From this perspective, we anticipate that, with increasing data size, participants need to scroll more, which will make the task more challenging for all conditions.
Precision is often required in scrolling, so increasing the data size may affect VR more than the Desktop.
On the other hand, it is possible to leverage the large display space in VR to alleviate this issue: we can use a large space to ensure all columns are visible without the need to scroll in VR. The user then can physically move in the space to navigate the large data tables. Such physical navigation has been found more effective than virtual navigation like panning or scrolling~\cite{ball2007move}. However, additional interaction supports are likely to be needed (e.g., shrinking the size of the table when it is grabbed) and should be tested in future work.
Our study did not include commercial data transformation tools on the desktop like Tableau Pre Builder and Trifacta. Those tools include many more features besides the basic data transformation operations, which may distract us from studying our intended main effects (e.g., WIMP vs. Gesture and Desktop vs. VR).
Adapting necessary features from those commercial desktop tools to VR is a promising next step, along with an ecological investigation of a complete system.
\added{Beyond the size of the datasets used, conducting user studies with a larger number of participants also should be tested. We expect this to lower the variance in several measures and further strengthen our hypothesis.}

\textbf{VR vs. AR.}
Our study was conducted in VR, as the VR hardware is more mature than AR headsets (e.g., higher resolution and larger field of view).
On the positive side, our designed gestures and implemented prototypes can be easily migrated from VR to AR thanks to the developing ecosystem, i.e., Unity3D.
We also believe our identified benefits in VR can be generalized to AR \added{or hybrid systems of AR and Desktop~\cite{wang2020towards}}, because the same as VR, AR also provides embodied interaction and large display space.
With AR hardware improving, it is important to investigate AR's unique challenges and benefits.
For example, observing the physical environment can improve people's spatial memory~\cite{luo2022should} and seamlessly switch between different computing environments~\cite{wang2020towards}.
We will systematically analyze and evaluate data transformation in AR in a future study once better AR headsets are released.

\section{Conclusion}
\label{sec:conclusion}
In this paper, we presented our prototypes to enable embedded interactions on the Desktop and embodied interactions in VR for data transformation.
We conducted a controlled user study to systematically evaluate the effect of the computing environments (Desktop vs. VR) and interaction methods (WIMP vs. Gesture).
We found initial evidences showing the benefits of using VR for data transformation:
VR had the potential to provide improved provenance over Desktop and demonstrated preliminary evidence of promoting strategic thinking.
VR also had a similar time performance compared to Desktop.
Additionally, participants found VR to be more engaging, and VR+Gesture provided the best overall experience.
Considering the iterative nature of data science, we foresee strong initiatives to combine immersive data transformation and visualization to enable a more complete \emph{Immersive Analytics} workflow, reducing the overhead of switching between different computing environments.
On the other hand, we did not find a better performance of Gesture over WIMP.
We still believe there is potential to improve Gestures, for example, by providing long-term training and real-time memory assistance. 
In summary, our results provide preliminary evidence that the large display space, spatial memory, and embodied navigation in VR are beneficial for high-level data transformation tasks.




\section*{Acknowledgments}
This work was supported in part by NSF I/UCRC CNS-1822080 via the NSF Center for Space, High-performance, and Resilient Computing (SHREC) and NSF grant III-2107328.

%








\bibliographystyle{IEEEtran}
\bibliography{main}

\begin{IEEEbiography}[{\vspace*{-5mm}\includegraphics[width=1in,height=1.1in,clip,keepaspectratio]{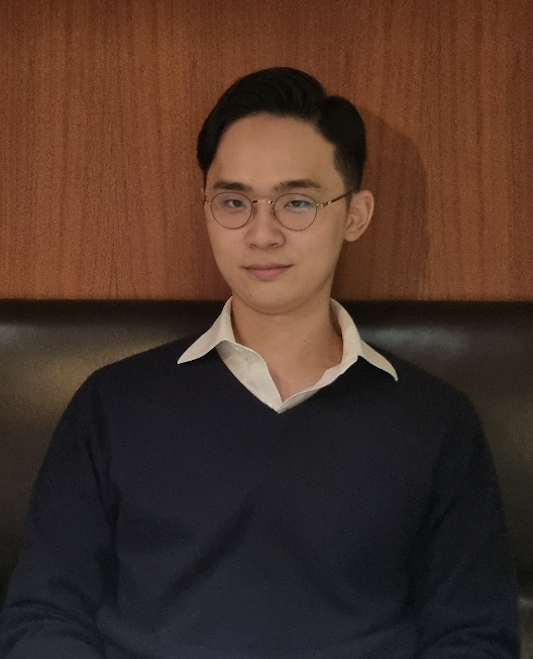}}]{Sungwon In} 
is a Ph.D. student at Virginia Tech. Before moving to Virginia Tech, he earned Bachelor's Degree in Computer Science from Indiana University, Bloomington, and he was an A.I. Software Engineer at FiniView. His research interests include the field of data transformation, data visualization, immersive analytics, and human-computer interaction, especially in 3D immersive environments (VR/AR).
\end{IEEEbiography}
\vspace{-13mm}

\begin{IEEEbiography}[{\vspace*{-5mm}\includegraphics[width=1in,height=1.1in,clip,keepaspectratio]{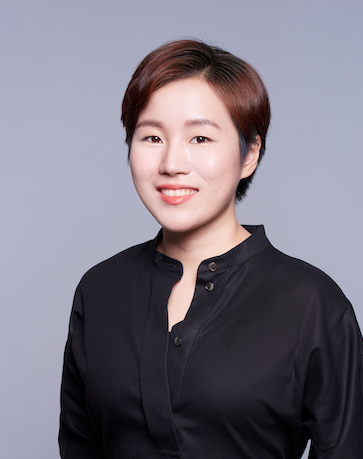}}]{Tica Lin}
is a Ph.D. student at the Visual Computing Group at Harvard University. Prior to joining Harvard, she was a Data Visualization Designer at Visa and a UX Developer at NBA 76ers. Her research interests include data visualization, immersive analytics, and human-computer interaction. In particular, she explores novel visualization and interaction design in Augmented Reality.
\end{IEEEbiography}
\vspace{-13mm}

\begin{IEEEbiography}[{\vspace*{-5mm}\includegraphics[width=1in,height=1.1in,clip,keepaspectratio]{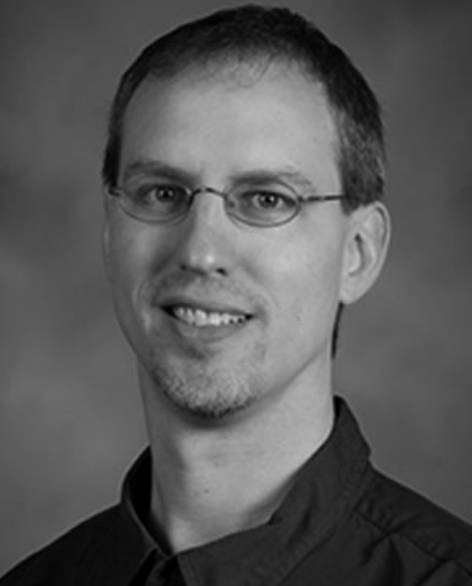}}]{Chris North}
is a professor of computer science with Virginia Tech. He is the associate director of the Sanghani Center for Artificial Intelligence and Data Analytics. His research
seeks to create effective methods for human-in-the-loop analytics of big data. His work falls in visual analytics, information visualization, human-computer interaction, large high-resolution display and interaction techniques, and visualization evaluation methods, with applied work in intelligence analysis, cyber security, bioinformatics, and GIS.
\end{IEEEbiography}
\vspace{-10mm}

\begin{IEEEbiography}[{\vspace*{-2mm}\includegraphics[width=1in,height=1.1in,clip,keepaspectratio]{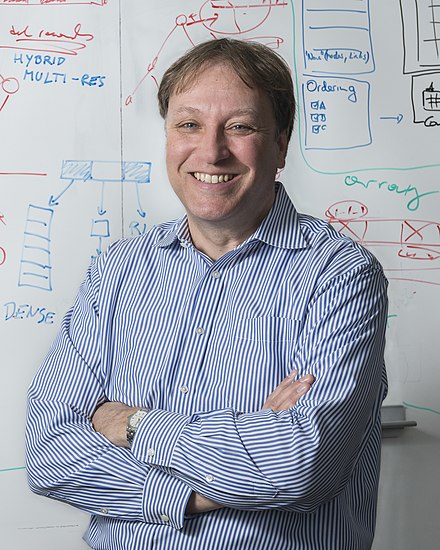}}]{Hanspeter Pfister}
is the Academic Dean of Computational Sciences and Engineering and An Wang Professor of Computer Science in the John A. Paulson School of Engineering and Applied Sciences at Harvard University. He has a Ph.D. in Computer Science from Stony Brook University and an MS degree in electrical engineering from ETH Zurich, Switzerland. His research in visual computing lies at the intersection of visualization, computer graphics, and computer vision and spans a wide range of topics, including biomedical visualization, image, and video analysis, machine learning, and data science.
\end{IEEEbiography}
\vspace{-10mm}

\begin{IEEEbiography}[{\vspace*{-5mm}\includegraphics[width=1in,height=1.1in,clip,keepaspectratio]{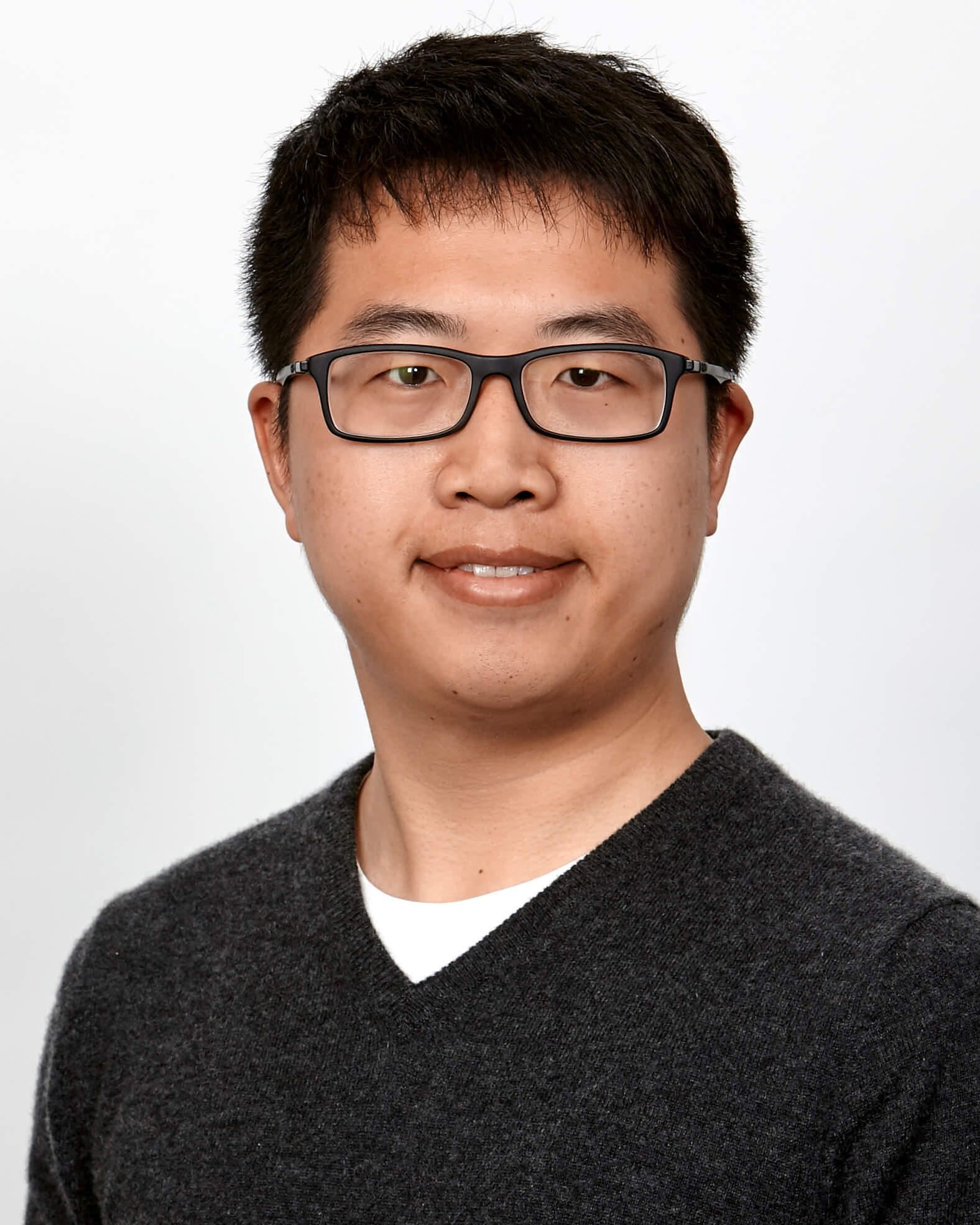}}]{Yalong Yang} is an Assistant Professor in the School of Interactive Computing, Georgia Institute of Technology.
He was an Assistant Professor at Virginia Tech, a Postdoctoral Fellow at Harvard University, and a Ph.D. student at Monash University, Australia. His research designs and evaluates interactive human-data interaction systems on both conventional 2D screens and in 3D immersive environments (VR/AR). 
\end{IEEEbiography}
\vfill

\end{document}